**Discipline and Resistance: The Construction of a "Digital Home" for TikTok Refugees on *Xiaohongshu***

XIONG Xiaoyu[1], PENG Yuting[2], KWONG Summer[3] and HUANG Anqi[4]

[1,2,3,4]The Chinese University of Hong Kong

**Abstract**

This study examines how "TikTok refugees" moved to *Xiaohongshu* after TikTok was about to be banned in the United States. It utilizes Foucault's idea of heterotopia to demonstrate how *Xiaohongshu* became a crisis space for cross-cultural discussions across the Great Firewall. Through Critical Discourse Analysis (CDA) of 586 user comments, the study reveals how Chinese and international users collaboratively constructed and contested a new online order through language negotiation, identity positioning, and playful platform policing. The findings highlight distinct discursive strategies between domestic and overseas users, reflecting both cultural resistance and adaptation. This research contributes to the understanding of digital migration, heterotopic spaces in social media, and emerging dynamics of cross-cultural discourse during geopolitical crises.

**Background**

The TikTok refugee phenomenon took place in a geopolitical anxiety context where TikTok risked a statewide ban unless it divested from its American operations by January 19th, according to a US government mandate. The U.S. government moved to ban TikTok due to national security concerns. Lawmakers argued that the app, owned by Beijing-based ByteDance, could enable Chinese authorities to influence content or access private user data. As a result, Congress passed a law requiring the platform to be sold to an approved buyer or face removal from American app stores (The New York Times, 2025). In response, American users scared of losing their platform flooded China's *Xiaohongshu* social media network, circumventing the Great Firewall, which typically divided internet ecosystems around the world (BBC, 2025). Within the two days, over 700,000 new users joined *Xiaohongshu*, and the US downloads of the app had reached a skyrocketing height of 200% year-over-year (Reuters, 2025). With the support of a single app architecture with no separate domestic and international versions, this astonishing movement briefly breached digital barriers, propelling *Xiaohongshu* to the top of US app stores and promoting a unique cross-firewall intercultural discourse.

Although there are several aspects being brought up in academia, namely digital migration and informal language learning (Liu et al., 2025) and cross-cultural communication (Zhang et al., 2025) that are discussed and researched. This project is particularly interested in investigating how Chinese users of *Xiaohongshu* create a welcoming space, how this online environment can become a heterotopia, defined by Foucault as a "counter-sites" and "a kind of effectively enacted utopia," (Foucault, 1984), for "TikTok refugees" from other countries, and how Chinese and international users modify this digital world within this heterotopia, demonstrating both resistance and discipline.

## Literature review

**Heterotopia: An Overview**

Michel Foucault coined the term "heterotopia," which means "other places," in 1966 and defined heterotopias as places that "draw us out of ourselves, where the erosion of our lives, time, and history takes place—spaces that torment and consume" (Foucault & Miskowiec, 1986). Foucault presented various instances of religious, artistic, sporting, and recreational venues, such as schools, honeymoon destinations, psychiatric institutions, prisons, cemeteries, theatres, libraries, museums, festivals, holiday resorts, and so on. What these locations have in common is their power to disturb the perceived continuity and routine of daily life.

In *Of Other Spaces* (1986), Michel Foucault introduces the concept of heterotopias, distinct spaces embedded within society yet governed by their own logic. These sites function as reflective surfaces, simultaneously mirroring and subverting the dominant norms of their cultural contexts. Foucault's six principles are not fixed classifications but flexible characteristics that manifest differently across time and space, shaping social dynamics in both material and symbolic ways. We draw on Lee and Wei's (2020) interpretation of these principles, particularly their application in *Social Media as Heterotopia*, which examines how digital platforms function as networked publics. Their framework provides a useful lens for analyzing *Xiaohongshu* as a heterotopia, as it bridges Foucault's theory with contemporary social media's unique spatial and temporal interventions. First, heterotopias have cultural specificity, which indicates each society develops unique heterotopias reflecting its values and norms. Second, heterotopia also demonstrates the process of evolving forms, which spaces transform across historical periods, adapting to cultural shifts. Third, they exhibit spatial layering, where a single heterotopia can contain multiple coexisting meanings, like a Persian carpet representing a miniature world within a room. Fourth, heterotopias process the attribute of temporal accumulation where they preserve and organize time, as seen in museums or libraries that archive historical periods. Fifth, each heteropia has controlled access, where entry is deliberately regulated through physical or symbolic boundaries, exemplified by gated communities. Lastly, functional contrast is an attribute where these spaces serve purposes distinct from ordinary spaces, like cemeteries addressing collective spiritual needs.

Traditional heterotopias have given way to more modern, functional spaces, with some scholars claiming that entire cities now serve as heterotopias (De Cauter & Dehaene, 2008). The emergence of digital technology has broadened the term, with cyberspace being presented as a new type of heterotopia (Rymarczuk and Derksen, 2014; Wark, 1993; Young, 1998). Within cyberspace, social media such as Facebook is also being identified as a digital heterotopia as it mirrors, distorts, and monetises reality itself (Rymarczuk, R., and Derksen, M. , 2014).

### *Xiaohongshu* As a Crisis Heterotopia

Foucault's (2008) conceptualization of heterotopias includes "crisis heterotopias" that emerge during societal disturbances to provide a temporary sanctuary or alternative order. As Foucault originally observed: "In primitive societies, there exists a form of heterotopia I call crisis heterotopias "privileged, sacred, or forbidden places reserved for individuals in a state of crisis relative to their social environment - adolescents, menstruating women, the elderly, etc." The quick shutdown of TikTok in the United States created precisely such a crisis scenario. Deprived of their habitual platform, US netizens unexpectedly migrated to *Xiaohongshu*, a Chinese social media app typically restricted by the Great Firewall. This phenomenon mirrors Foucault's framework while demonstrating its digital evolution: where crisis heterotopias historically provided physical refuge for biological transitions, we observed digital platforms now offer virtual asylum during geopolitical-tech disruptions.

Notably, as Foucault noted, "These crisis heterotopias are disappearing today, being replaced by heterotopias of deviation." *Xiaohongshu* exhibits this dual yet paradoxical function, a "deviation heterotopia" through its algorithmic governance and content controls, while simultaneously serving as a "crisis heterotopia" for displaced users. Its single-app architecture (lacking separate regional versions) creates what Foucault might call a "heterotopia without geographical markers," temporarily suspending normal digital boundaries during the platform migration crisis.

**Refugee Studies**

The current definition of a refugee varies, and no single definition exists. However, a refugee could mean 'in essence, a person who has crossed an international frontier because of a well-founded fear of persecution' (Shacknove, 1985, p.274). These people are suffering due to their country governments' inability to fulfil their needs. They have no alternative but to seek international assistance (Shacknove, 1985).

Discussions on refugee issues often focus on refugees moving to Western countries and the associated online media (especially social media) (Siapera, 2004; Barisione et al., 2019). Regarding refugees from Western nations, governments require refugees to complement their national labour force, but they are reluctant to provide these refugees with social welfare and citizenship rights. This situation leads to refugees' disadvantaged position (Siapera, 2004). Siapera (2004) considers that social media are vital to refugees' survival. They may encounter discrimination offline from governmental or non-governmental organizations, requiring them to access information more effectively through the internet. Through the internet, refugees can share group stories, obtain legal advice, and obtain helpful information from experts and the general public. Simultaneously, refugees can gain sympathy from citizens of migrant countries through social media. For example, in 2015, internet users in Germany and the UK expressed their sympathy for the refugee community by using the hashtag #RefugeesWelcome on Twitter (Barisione et al., 2019). However, language barriers may cause issues with their online enquiry and participation. Social media platforms in countries of immigration are still dominated by the local language (Jimenez-Andres, 2021). Alencar (2020) argues that these refugees are eager to assimilate into the host culture and thus engage in active language and cultural learning.

Apart from access to valuable information online, the environment does not appear to be entirely positive (Filibeli & Ertuna, 2021; Ekman, 2018). The anonymity of the internet allows users to voice extreme emotions and radical ideas freely. The refugee community is exposed to online hate speech. For instance, Turkish hate speech against Syrian refugees on Facebook. Turkish people employ hashtags such as #SyriansGoAway (Filibeli & Ertuna, 2021). Filibeli & Ertuna (2021) illustrate that through the use of these hashtags, Turkish people express their negative feelings towards refugees. In terms of specific content, they create fake news to enhance the negative image of these refugees. Ekman (2018) notes that these nationals rationalise discrimination against refugees as self-defence. Refugees are a violent and existential threat to their lives. The creation of all this content reflects the emotional drive between the different groups (Barisione et al., 2019).

The term 'digital refugee' has appeared, and it is mainly used to describe people who cannot adapt to the digital environment and fully utilize digital technologies to access and process information effectively. Nevertheless, those who have moved to social media platforms in other countries due to national restrictions (Combes, 2021). American users who have turned to *Xiaohongshu* after the US government banned TikTok can be considered 'digital refugees' to some extent. Similar to traditional or offline refugees, they are unable to access TikTok in their own countries and cross online borders to China's unique social media platform. These 'TikTok refugees' shed light on the virtual asylum behaviors resulting from social media blocking, offering new perspectives on transnational migration in the digital age. Additionally, existing research focuses on traditional refugees instead of online migrants. These TikTok refugees may refer to traditional refugees when communicating on host country platforms, with host country citizens emphasising emotions, language, and cultural integration.

**Social Media and Cross-culture Communication**

Social media has enabled people from various backgrounds across the world to interact and communicate without the constraints of time and distance (Adenubi &

Adegboyega, 2020; Sawyer & Chen, 2012). In the context of globalization, social media has the capacity to facilitate cross-cultural exchanges between different populations. These exchanges have also shaped the globalization process to a certain extent. However, such cross-cultural interactions present both opportunities and challenges (Adenubi & Adegboyega, 2020; Balogun& Aruoture, 2024).

Early studies on cross-cultural research focused on international social media platforms like Facebook, YouTube, and Twitter (Adenubi & Adegboyega, 2020; Sawyer & Chen, 2012; Balogun & Aruoture, 2024). Regarding opportunities, using social media has increased the opportunities for marginalized cultures to be displayed and facilitated cultural communication. Specifically, on these international social media platforms, Nigerian users post their marginalized indigenous cultures to the platforms, thus allowing these practices to be documented and shared (Balogun & Aruoture, 2024). Social media can also help those who wish to learn about other countries' cultures to achieve their goals. Sawyer & Chen (2012) find that some non-Americans use Facebook to ask American users for guidance on local slang, thus allowing them to better understand American culture.

However, these social media dominated by Western culture may promote cultural homogenization or the infiltration of cultural imperialism (Balogun & Aruoture, 2024). Since these well-known international social media platforms are created by Westerners, their algorithms prefer to recommend Western culture, especially American culture. This situation further contributes to the Western culture's hegemony (Balogun & Aruoture, 2024). Tang & Chan point out that in the event of prolonged dominance of Western culture, it may erode the other regions' cultures, resulting in cultural homogenization. Users are affected by behaviors and cultural mindsets (Qiu & Leung, 2013). According to Tang & Chan (2020), some Malaysians who use these social media platforms begin to practice Western culture and abandon their own culture, reflecting global cultural homogenization.

*Xiaohongshu*, a famous domestic social media platform in China, has a recommendation algorithm (collectivism) and a censorship system based on the culture of Chinese people (Liu, 2024). Therefore, cross-cultural communication on this platform may have different outcomes than in the West. Meanwhile, after the emergence of TikTok Refugees, *Xiaohongshu* is seen as possessing the potential to shape a dynamic space for cross-cultural exchange. Due to the use of firewalls and different social media platforms, this phenomenon is unprecedented. On the platform, Chinese users can spontaneously discuss different or similar life and cultural experiences with newly arrived foreign users (Liu et al., 2025). Therefore, the study of TikTok refugees on the *Xiaohongshu* platform is significant because it fills in the absence of new forms of digital refugees and provides new perspectives on cross-cultural research between China and America. Consequently, based on the significance of the literature review above, the research question of this study is: In the context of TikTok refugees flooding into *Xiaohongshu,* how is *Xiaohongshu* constructed as a heterotopia within the digital sphere?

## Method

This study employs Critical Discourse Analysis (CDA), grounded in Fairclough's (1992) three-dimensional model, to explore user interactions surrounding the migration of "TikTok refugees" to *Xiaohongshu*. We collected over 800 TikTok refugee-related posts from *Xiaohongshu* between January 13 and February 8, 2025, using a custom web scraper (https://github.com/NanmiCoder/MediaCrawler). Each post included metadata such as content, tags, user location, and media links; comments were linked by note_id and included text, timestamps, likes, and user information. Through systematic coding and qualitative analysis of selected user comments, this CDA approach examines textual features, interactional patterns, and broader social practices, emphasizing comparative analysis between domestic and overseas users.

Figure 1. Fairclough's 3D model of critical discourse analysis. (1st dimension = Inner layer, 2nd dimension = middle layer, 3rd dimension = outer layer).

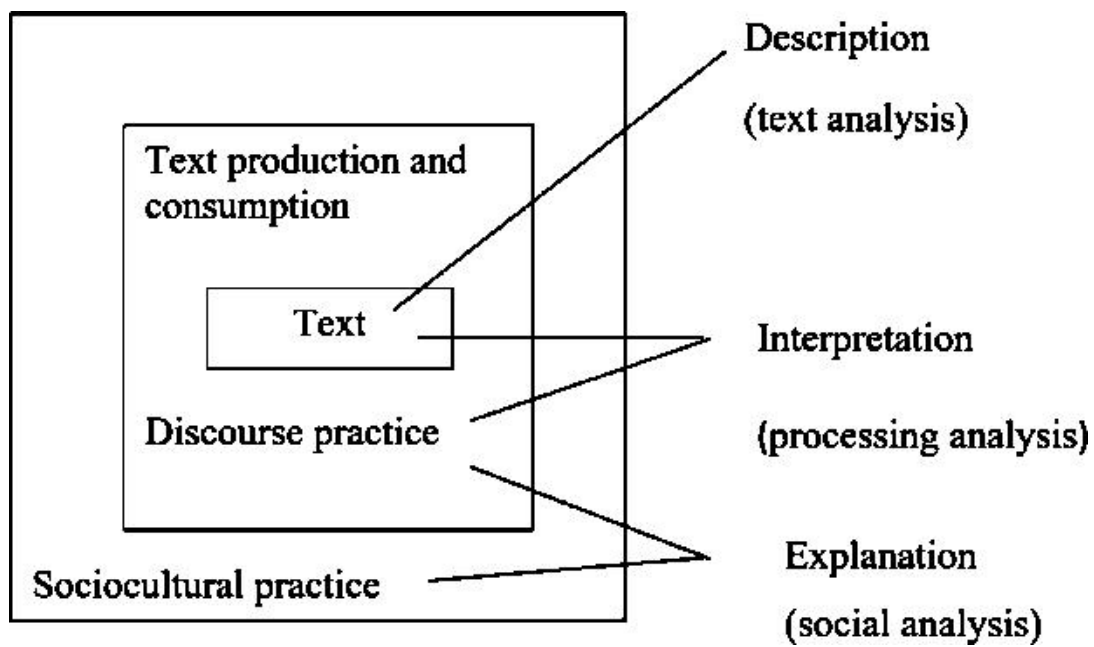

**Codebook Construction and Comment Selection**

Drawing on prior CDA-informed studies on digital and cross-cultural discourse (Wark, 1994; Rymarczuk & Derksen, 2014; Jones et al., 2015; Chen, 2018; Zhu et al., 2024), we first developed an initial set of **ten focus codes** capturing key discursive mechanisms such as language mixing, identity labeling, intertextuality, ideological alignment, and playful platform regulation. These codes were theoretically aligned with Fairclough's model and categorized accordingly. The theoretical origins and analytical scope of each code are detailed in **Table 1**, which maps each focus code to its primary discourse layer and the supporting literature.

Table 1. Theoretical origins of focus codes used in CDA

| Focus Code | Theoretical Basis / Source | Discursive Layer |
|---|---|---|
| **Language Mixing / Translanguaging** | Jones et al. (2015): Identity negotiation via bilingual/multilingual YouTube comments | Textual Features |
| **Identity Labels / Self-positioning** | Jones et al. (2015); Chen (2018): Cultural boundary marking and identity positioning in social platforms | Textual Features & Social Practice |
| **Attitude Markers** | Fairclough (1992); Ludemann (2021): Stance, modality, and evaluative expressions in cultural discourse | Textual Features |
| **Cultural/Platform Intertextuality** | Chen (2018); Fairclough (1992): Intertextuality and heterotopic cultural references | Textual + Discursive Practice |
| **Interaction Type** | Jones et al. (2015); Ludemann (2021): Communicative functions, collaborative learning, digital interaction routines | Discursive Practice |
| **Humor & Playful Policing** | Chen (2018); Fairclough (1992): Use of humor and satire as community regulation in heterotopic spaces | Discursive + Social Practice |
| **Role Reversal / Identity Shifting** | Ludemann (2021): Reversing knowledge roles in translingual encounters | Discursive Practice & Social Practice |
| **Cultural Identity & Negotiation** | Chen (2018); Rymarczuk & Derksen (2014): Cultural heterogeneity, identity change through online discourse | Social Practice |
| **Ideology / Platform Politics** | Fairclough (1992); Rymarczuk & Derksen (2014): Power, censorship, digital governance as ideological manifestations | Social Practice |
| **Globalization Awareness** | Wark (1994); Rymarczuk & Derksen (2014): Transnational imaginaries, global user subjectivity in cyberspace | Social Practice |

To apply this coding framework meaningfully, we then curated a subset of user comments from our *Xiaohongshu* dataset through a two-step filtering process. First, we retained only those comments that semantically matched at least one of our focus codes, based on keyword patterns and cross-cultural discursive cues (e.g., "TikTok," "laowai," "cat tax," "freedom"). Second, we excluded trivial or non-analytic comments by applying a minimal length threshold of 8 characters. This resulted in a high-quality, semantically dense corpus of 586 comments suitable for critical discourse analysis.

Following this filtering, we conducted iterative rounds of trial coding on a subsample to refine our definitions, ensure thematic clarity, and account for overlapping discursive functions. The final version of the codebook retained ten core codes, each with a precise definition, theoretical anchor, and representative examples drawn directly from the filtered dataset, see Appendix 1. This refinement process reflects Fairclough's emphasis on the **layered nature of discourse,** wherein even a single utterance may operate simultaneously at textual, interactional, and ideological levels.

**Coding Procedure and Analytical Strategy**

Following the finalization of our codebook and selection of relevant comments, we conducted a structured coding procedure to ensure analytical rigor and reliability. Three independent coders were trained based on the finalized codebook to code the full dataset of 586 comments individually. After completing the initial independent coding, the coders convened to review discrepancies and resolve any disagreements through consensus discussion, resulting in a refined and unified dataset suitable for subsequent analysis. This iterative procedure helped guarantee internal consistency and reliability of our coding results (Krippendorff, 2018).

Given Fairclough's (1992) emphasis on discourse as a form of social practice and identity construction, we further categorized our dataset based on geographic origin (China vs. overseas IP addresses). This decision was informed both theoretically and contextually. Theoretically, Fairclough (1992) underscores that discourse patterns can vary significantly depending on social identities and positions, suggesting users from different cultural and geographic backgrounds may display distinct discursive practices, identity negotiations, and ideological alignments. Empirically, considering the specific context of our research—the migration of "TikTok refugees" to *Xiaohongshu*—there is a strong rationale for expecting variations in discursive practices among users of different geographical origins. Users from overseas IPs (primarily associated with the migrating TikTok community) might adopt different discursive strategies compared to domestic Chinese users who are already integrated into *Xiaohongshu*'s local social norms and practices. These differences reflect the broader geopolitical contexts, cultural boundaries, and identity negotiations inherent in this digital migration phenomenon (Jones et al., 2015; Rymarczuk & Derksen, 2014).

After the coding procedure and geographic classification, descriptive statistics were calculated for the frequency distributions of each focus code within the overall dataset, as well as separately for China and overseas IP datasets, respectively (see Table 2). Additionally, to capture Fairclough's notion of multilayered discursive embedding, we conducted co-occurrence analyses, visualized through heatmaps (Figures 1-3). This allowed us to objectively identify patterns of overlapping discourse categories, revealing how multiple discursive practices frequently co-occur within single user comments.

## Results

**Frequency Distribution of Focus Codes**

Table 2 Frequency Distribution of Focus Codes

| Focus Code | Overall (n = 586) | China IP (n = 403) | Overseas IP (n = 183) |
|---|---|---|---|
| Language Mixing / Translanguaging | 148 (25.26%) | 109 (27.05%) | 39 (21.31%) |
| Identity Labels / Self-positioning | 162 (27.65%) | 117 (29.03%) | 45 (24.59%) |
| Attitude Markers | 344 (58.7%) | 245 (60.79%) | 99 (54.1%) |
| Cultural/Platform Intertextuality | 199 (33.96%) | 143 (35.48%) | 56 (30.6%) |
| Interaction Type | 197 (33.62%) | 153 (37.97%) | 44 (24.04%) |
| Humor & Playful Policing | 89 (15.19%) | 67 (16.63%) | 22 (12.02%) |
| Role Reversal / Identity Shifting | 26 (4.44%) | 15 (3.72%) | 11 (6.01%) |
| Cultural Identity & Negotiation | 145 (24.74%) | 106 (26.3%) | 39 (21.31%) |
| Ideology / Platform Politics | 100 (17.06%) | 61 (15.14%) | 39 (21.31%) |
| Globalization Awareness | 145 (24.74%) | 111 (27.54%) | 34 (18.58%) |

The descriptive statistical analysis (see Table 2) reveals distinct patterns of focus code usage across the entire dataset (n = 586), as well as notable differences between China (n = 403) and overseas IP users (n = 183). Overall, the most prominent focus code was "Attitude Markers" (58.7%, n = 344), indicating that a significant majority of comments expressed explicit emotional evaluations such as welcome, praise, or irony. Additionally, "Interaction Type" (34.0%, n = 199) and "Ideology / Platform Politics" (33.6%, n = 197) were frequently observed, emphasizing active user interactions and strong ideological engagement related to platform governance and cultural boundaries.

A comparative analysis further reveals nuanced differences between China and overseas IP users. Among China-based users, besides "Attitude Markers" (60.8%, n = 245), there was notably high engagement with "Ideology / Platform Politics" (38.0%, n = 153), reflecting users' critical or evaluative stances toward the platform's cultural and ideological environment. In contrast, overseas IP users placed relatively stronger emphasis on "Cultural Identity & Negotiation" (24.6%, n = 45) and "Interaction Type" (30.6%, n = 56). This suggests overseas users were more actively engaged in negotiating their identities within the new cultural context, and they were explicitly interactive in their approach to platform engagement.

Overall, these clear distinctions in focus code usage reflect deeper variations in discursive priorities between domestic and international user groups, highlighting how geographic and cultural differences shape users' online discourse practices and identity negotiations (Fairclough, 1992; Chen, 2018).

**Co-occurrence Patterns of Focus Codes**

The co-occurrence heatmaps (Figures 2–4) further illustrate how different discourse categories interweave within individual comments, directly reflecting Fairclough's (1992) conceptualization of discourse as multilayered and embedded simultaneously across textual, interactional, and social dimensions. In the overall dataset, the strong co-occurrence between "Attitude Markers" and "Cultural/Platform Intertextuality" (n = 143) indicates that users' emotional or evaluative stances often rely on intertextual references that bridge multiple cultural and digital platforms, thereby connecting micro-level textual expressions with broader discursive practices. Similarly, the substantial co-occurrence observed between "Globalization Awareness" and "Attitude Markers" (n = 104) illustrates how global consciousness and reflections on digital migration frequently occur alongside explicit emotional attitudes, underscoring the nested relationship between individual evaluations and macro-social realities within user discourse.

Figure 2 Co-occurrence Matrix of Focus Codes

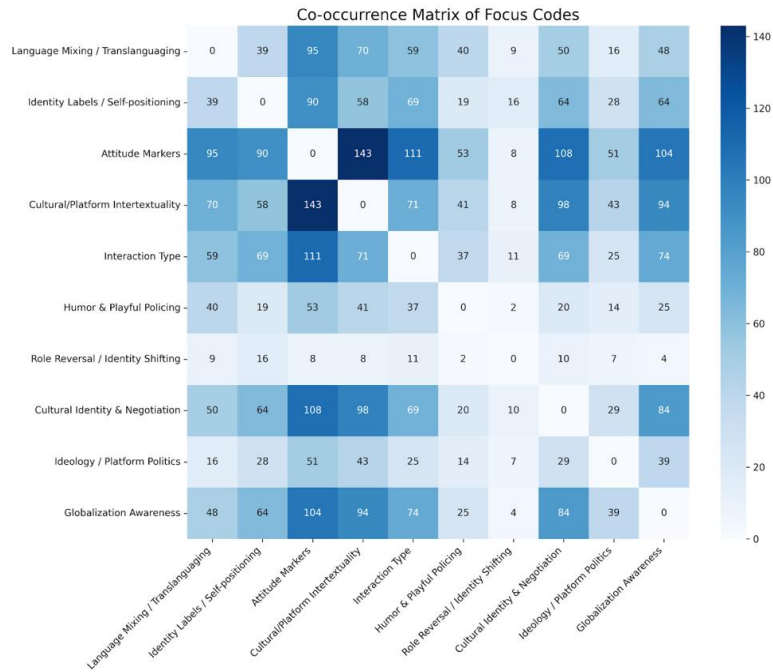

Figure 3 Co-occurrence Matrix of Focus Codes (China lP)

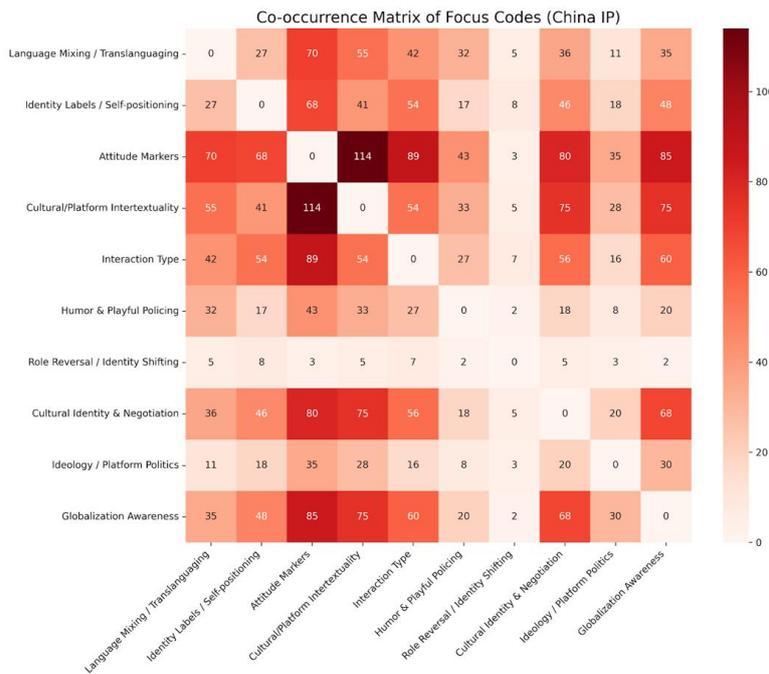

Figure 4 Co-occurrence Matrix of Focus Codes (Overseas lP)

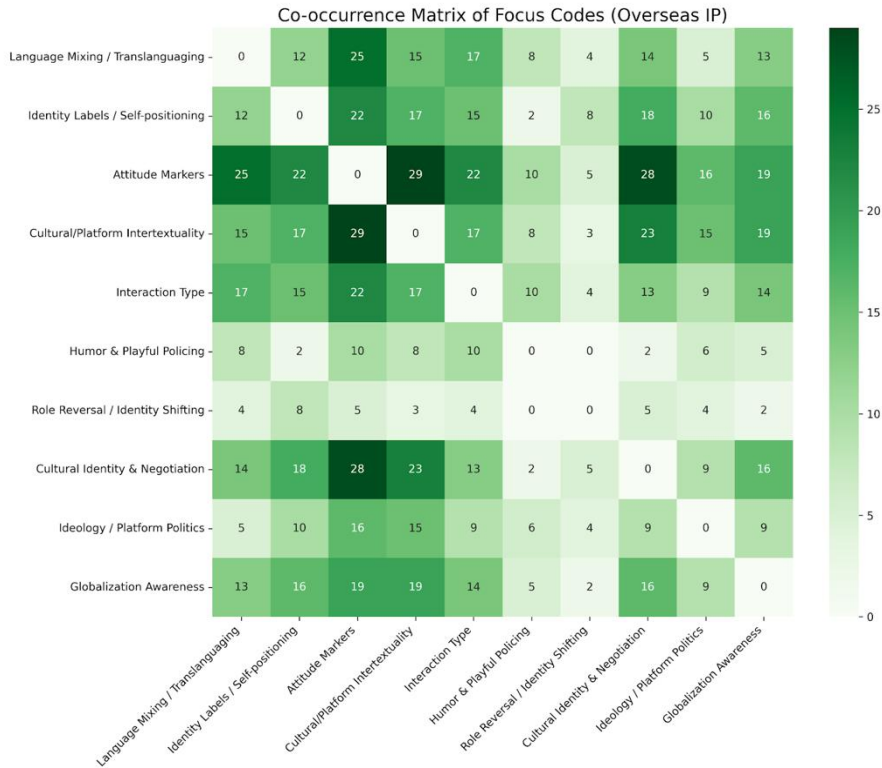

The patterns among China IP users (Figure 3) further demonstrate such multilayered embedding. Specifically, the frequent co-occurrence of "Attitude Markers" with "Ideology / Platform Politics" highlights how emotional expressions are intimately tied to ideological positioning, reflecting users' conscious negotiations of platform norms and broader cultural and ideological environments. Conversely, overseas IP users (Figure 4) prominently combine "Interaction Type" with "Cultural Identity & Negotiation," emphasizing how active interactive practices intersect closely with cultural identity formation and negotiation. This distinct pattern illustrates overseas users' discursive practices as an explicit attempt to navigate and redefine identity within the new digital and cultural setting of *Xiaohongshu*.

These findings further substantiate Fairclough's (1992) emphasis on the inherent complexity and multi-dimensional nature of discourse by demonstrating empirically how even seemingly simple user comments simultaneously embody textual nuances, interactional dynamics, and deeper social and ideological practices.

## Discussion

**The Construction of Heterotopia in *Xiaohongshu***

As a crisis heterotopia formed during the "TikTok ban" crisis, *Xiaohongshu* underwent two stages of transformation: Stage 1: Overcoming technological and linguistic barriers; Stage 2: Re-establishing new rules for cross-cultural communication on *Xiaohongshu*.

*"So why do we have to translate into English when we go to waiwang (外网), but when they come here, we're still the ones speaking English? Can you speak Chinese?"* (Translation in English version, from mainland China, 2025-01-14)

The term "waiwang" (外网) is used by Chinese internet users to refer to overseas internet platforms. As previously mentioned, the "Great Firewall of China" implements various types of censorship and content filtering to regulate internet traffic in China, playing a key role in maintaining cybersecurity and controlling public opinion (Ensafi et al., 2015). Under this system, it is challenging for Chinese internet users to access certain international websites or applications (e.g., Instagram, Facebook). At the same time, overseas internet users also face difficulties accessing information from within the Chinese internet. *Xiaohongshu* has served as an interstitial ground in this context, becoming one of the few platforms exempt from the full enforcement of the "Great Firewall" (Ensafi et al., 2015). Consequently, it has acted as a bridge connecting Chinese internet users with overseas users. By transcending the physical limitations of digital borders, *Xiaohongshu* grants users the ability to "cross boundaries," enabling foreign users to access Chinese social platforms and exemplifying the heterotopian principle of "spatial layering" (Rymarczuk & Derksen, 2014).

"*2 元帮起中文名，20 元看八字，98 元改命。*

*￥2 to help you get a Chinese name, ￥20 to calculate hexagrams, ￥98 to change your destiny.*" (From mainland China, 2025-01-15; the above text is in Chinese, followed by its corresponding English translation)

To address the language barrier encountered by overseas users entering *Xiaohongshu* (as the platform had not yet introduced an online translation feature at that time), various cross-linguistic communication methods emerged through negotiation between local and overseas users. In our findings, "Language Mixing/Translanguaging" (25.26%) primarily manifested as users including translations of their content in their native language. This display of cultural inclusivity and friendliness helped to diminish cultural divides, creating a "translingual heterotopia" (Jones et al., 2015), where users utilized two languages in their interactions, often accompanied by humor, to achieve identity negotiation "in a tactical and poetic way" (Chen, 2018, p. 273).

Simultaneously, within the formation of this heterotopia, TikTok refugees experienced a breakdown of previously established platform logics from TikTok. In terms of cross-cultural translation, this often led to "semantic inconsistencies," resulting in discrepancies in meaning. To address this, Sperber (1994) proposed the method of "decentering," which involves iterative comparison and revision between the two languages until a culturally relevant and similar instrument is validated in both. During this cross-cultural exchange, TikTok refugees found that the linguistic expressions used on overseas platforms (e.g., "lol," "fyp") became ineffective on *Xiaohongshu*. They were thus compelled to adapt to localized rules (e.g., popular Chinese internet expressions such as "xsl" or "nb"), reflecting the heterotopian principle of "functional contrast" (Foucault & Miskowiec, 1986).

Throughout this process of cross-cultural communication, both local Chinese internet users and overseas users simultaneously reconstructed their cultural identities. For local users, "attitude markers" (60.79%) were employed to express either their welcome or rejection of the "refugees," while emphasizing *Xiaohongshu*'s platform rules to reaffirm their "local" identity (Fairclough, 1992). For TikTok refugees, "cultural identity & negotiation" (21.31%) and "interaction type" (30.6%) were utilized to establish a fluid identity that resonates with the compensatory-subversive capacity of heterotopias (Lee & Wei, 2020). For instance, they would adopt phrases such as "try world citizens instead of refugees" (from USA, 2025-01-29) to redefine their identity and mitigate the cultural disadvantage associated with being labelled as refugees.

**Power Dynamics: Discipline and Resistance**

While forming this crisis heterotopia, a subtle power dynamic of discipline and resistance existed between local users and TikTok refugees. During this process, *Xiaohongshu* serves as "a heterotopic space where netizens contest societal norms and challenge the hegemony of identities" (Wulandari et al., 2023, p. 109).

*"This is really great. It's been a long time since I've seen such a deep post on Xiaohongshu. Love it. The fact that they call themselves 'refugees' is actually a reflection of cultural hegemony, including how everyone cooperates by speaking English, which is*

*essentially a form of linguistic hegemony." (Translation in English version, from mainland China, 2025-01-14)*

As gatekeepers, local users strengthened their cultural identity during this cross-cultural exchange through "Humor & Playful Policing" (16.63%) and "Ideology / Platform Politics" (15.14%). For instance, the humorous strategy of "cat tax" became one of the most common tactics in this cross-cultural interaction, requiring overseas users to post cute photos of their pets as a "key" to access this heterotopia. Within this discourse, the term "cat tax" not only reflects Chinese users' playful critique of U.S. taxation culture but also serves as a measure to reinforce local norms, embodying the heterotopian principle of "controlled access" (Foucault & Miskowiec, 1986). Additionally, local users heightened their focus on content ideology by critiquing Western cultural hegemony (e.g., linguistic hegemony), thereby safeguarding *Xiaohongshu*'s local identity and maintaining their cultural dominance on this platform (Tang & Chan, 2020).

*"We gotta keep in mind that Americans, us, are their guests. We should be respectful and follow their rules. 我们要记住，作为美国人，我们是他们的客人。我们应该尊重他们，遵守他们的规则." (From USA, 2025-01-15)*

*"I never really liked the TikTok community. I am an American, and I will say that I do not like any US social media app. It's all just negative stuff and a toxic community. I would very much like to stay on here if I can." (From USA, 2025-01-21)*

Compared to local users, TikTok refugees, operating as "outsiders," adopted strategic compromises and resistance tactics. Through "Role Reversal/Identity Shifting" (6.01%), they adapted to the platform's rules by engaging in perspective-taking within the Chinese cultural context, attempting to reconstruct their identities (Ludemann, 2021) while creating a low-conflict, safe communication environment. Moreover, they combined "Globalization Awareness" (18.58%) with "Cultural/Platform Intertextuality" (30.6%), overcoming stereotypes and employing subtle approaches to dissolve cultural biases. In doing so, they transformed *Xiaohongshu* into a transnational cultural exchange laboratory (Wark, 1994).

**Temporary and Structural Dilemmas of Heterotopia**

The construction of this crisis heterotopia was primarily triggered by the sudden geopolitical event of the "TikTok ban". As a result, it embodies both temporary and structural dilemmas, reflecting not only the "normal" spatial network but also the complexities of social, cultural, and political arrangements (Dalton, 2014).

*"After a while, I opened Xiaohongshu and found that the videos were almost entirely by Chinese users, with barely any foreigners left—then it felt like the sky was falling." (Translation in English version, from mainland China, 2025-01-22)*

*"Out of the 117 foreign accounts I was following, 21 have been banned." (Translation in English version, from mainland China, 2025-01-21)*

For TikTok refugees, *Xiaohongshu* can only serve as a "short-term shelter." While overseas user migration momentarily breaks digital borders, it remains constrained by the "temporal accumulation" (Foucault & Miskowiec, 1986) of censorship under the "Great Firewall", making the operation of this heterotopia unsustainable. The platform's ability to retain overseas users is further limited by the cultural adaptation required of overseas users and the dominance of the Chinese language on *Xiaohongshu*. Additionally, *Xiaohongshu*'s recommendation algorithms gradually "tame" refugee content (e.g., limiting the reach of English-language posts and promoting homogenized content), eroding the platform's "compensatory-subversive capacity" as a heterotopia through structural governance (Lee & Wei, 2020).

> *"Thank you for clarifying the difference between the social credit system and the blacklist. That's definitely super eye-opening. Do you have any good recommendations for sources to read more? 感谢您澄清社会信用和黑名单之间的区别，这绝对是超级大开眼界。您对于阅读更多内容有什么好的建议吗？" (From USA, 2025-01-19)*

*"XHS, or Chinese censorship in general, is still very strict. Lots of content you can post on TikTok isn't allowed in China." (Translation in English version, from Singapore, 2025-01-13)*

*"Happy New Year, beautiful girl! Come to China to look around, have fun, and buy a Huawei phone while you're at it—problem solved!" (Translation in English version, from mainland China, 2025-01-31)*

As a bridge for cross-cultural communication during this event, *Xiaohongshu* has indeed fostered diversity (e.g., by facilitating mutual explanations of stereotypes between Chinese and overseas users). However, at the same time, the dominance of local users in controlling discursive power on the platform has reinforced local cultural hegemony, exposing the internal contradictions of the heterotopia (Tang & Chan, 2020).

Moreover, while the heterotopian principle of "controlled access" provides a "safety net" that accommodates resistance while preventing subversion (e.g., through the censorship of sensitive terms), it has also enabled local users to expand their subjectivity and autonomy within the platform (Chen, 2018). However, this same mechanism has also become a breeding ground for online populism, ultimately hindering further cross-cultural exchange.

The TikTok ban and the ensuing migration of users to *Xiaohongshu* reveal how digital platforms can temporarily function as crisis heterotopias, as theorized by Foucault. In response to geopolitical pressure, *Xiaohongshu* briefly became a counter-site where digital borders collapsed and unexpected cross-cultural encounters emerged. This study explored how domestic *Xiaohongshu* users fostered a welcoming environment for incoming TikTok refugees, how this platform evolved into a space of both resistance and discipline, and how users modified this digital heterotopia through daily interactions. Using Critical Discourse Analysis of user comments, we analyzed how such a crisis event reshaped online space and prompted the re-negotiation of cultural boundaries.

The findings contribute to research on cross-cultural communication and digital migration by showing how geopolitical disruptions can give rise to temporary yet meaningful intercultural spaces. The case of TikTok refugees on *Xiaohongshu* illustrates new dimensions of digital migration, distinct from traditional refugee movements. It also highlights how online platforms like *Xiaohongshu* serve not only as sites of cultural interaction but also as arenas where linguistic, social, and ideological differences are negotiated and managed. While existing scholarship often focuses on refugees entering Western countries, this study extends the conversation by examining virtual displacement and integration processes in non-Western digital contexts. It offers new insights into how marginalized or displaced digital communities navigate foreign cultural environments, negotiate language barriers, and engage in identity reconstruction under sudden geopolitical shocks.

This study shows how *Xiaohongshu* evolved into a crisis heterotopia that helped TikTok refugees cope with sudden digital displacement. Our findings reveal that local users leveraged humor and playful regulation to protect platform norms and resist external cultural hegemony, effectively using *Xiaohongshu* as a channel for cultural exportation. At the same time, TikTok refugees adapted by repositioning their identities, building global awareness, and engaging in cultural translation. However, the cross-cultural interactions also exposed the platform's structural contradictions: algorithmic governance restricted content diversity, and the dominance of domestic users risked reinforcing cultural hegemony rather than sustaining open dialogue. Moreover, this study only focuses on *Xiaohongshu*, and potential algorithmic biases could have influenced sample representation. Future research should investigate the longer-term evolution of interactions between TikTok refugees and *Xiaohongshu* users to assess the sustainability of cross-cultural engagement within crisis heterotopias.

**Conclusion**

The TikTok ban and the move of users to *Xiaohongshu* show that digital platforms can sometimes act as crisis heterotopias, as Foucault said. Because of geopolitical pressure, *Xiaohongshu* quickly became a counter-site where digital borders fell and new cross-cultural meetings happened. This study looked at how *Xiaohongshu* users in China created a welcoming space for TikTok refugees, how the platform became a place of both resistance and discipline, and how users changed this digital space through daily actions. We used Critical Discourse Analysis of user comments to study how this crisis changed online spaces and made users rework cultural boundaries.

The findings help research on cross-cultural communication and digital migration by showing that geopolitical problems can create short but important spaces for different cultures to meet. The case of TikTok refugees on *Xiaohongshu* shows new sides of digital migration that are different from normal refugee stories. It also shows that platforms like *Xiaohongshu* are not only places for cultural meetings but also places where people work through language, social, and idea differences. Most studies look at refugees moving into Western countries, but this study talks about online moves and how users join and adapt in non-Western spaces. It gives new ideas about how online groups facing sudden political changes deal with language, culture, and identity challenges.

De Cauter, L., & Dehaene, M. (2008). The space of play: Towards a general theory of heterotopia. In M. Dehaene & L. De Cauter (Eds.), Heterotopia and the city (pp. 99–114). Routledge.

Dehaene, M., & De Cauter, L. (2008). Heterotopia and the city: Public space in a postcivil society. Routledge.

Dehaene, M., & De Cauter, L. (2008). Heterotopia in a postcivil society. In Heterotopia and the City (pp. 15–22). Routledge.

Ekman, M. (2018). Anti-refugee mobilization in social media: The case of soldiers of Odin. Social Media + Society, 4(1), 2056305118764431.

Ensafi, R., Winter, P., Mueen, A., & Crandall, J. R. (2015). Analyzing the Great Firewall of China over space and time. Proceedings on Privacy Enhancing Technologies.

Fairclough, N. (1992). Discourse and social change. Polity Press.

Filibeli, T. E., & Ertuna, C. (2021). Sarcasm beyond hate speech: Facebook comments on Syrian refugees in Turkey. International Journal of Communication, 15(2021), 2236–2259.

Foucault, M., & Miskowiec, J. (1986). Of other spaces. Diacritics, 16(1), 22–27.

Jimenez-Andres, M. (2021). Refugee access to information in online and offline environments: Results from focus group discussions. Fitispos International Journal, 8(1), 79–95.

Jones, R. H., Chik, A., & Hafner, C. A. (2015). Discourse and digital practices: Doing discourse analysis in the digital age (1st ed.). Routledge. https://doi.org/10.4324/9781315726465